\documentclass[twoside,twocolumn,9pt]{article}
\usepackage{extsizes}
\usepackage[super,sort&compress,comma]{natbib} 
\usepackage[version=3]{mhchem}
\usepackage[left=1.5cm, right=1.5cm, top=1.785cm, bottom=2.0cm]{geometry}
\usepackage{balance}
\usepackage{mathptmx}
\usepackage{sectsty}
\usepackage{graphicx} 
\usepackage{lastpage}
\usepackage[format=plain,justification=justified,singlelinecheck=false,font={stretch=1.125,small,sf},labelfont=bf,labelsep=space]{caption}
\usepackage{float}
\usepackage{fancyhdr}
\usepackage{fnpos}
\usepackage[english]{babel}
\addto{\captionsenglish}{%
  
}
\usepackage{multirow}
\usepackage{array}
\usepackage{droidsans}
\usepackage{charter}
\usepackage[T1]{fontenc}
\usepackage[usenames,dvipsnames]{xcolor}
\usepackage{setspace}
\usepackage[compact]{titlesec}
\usepackage{hyperref}
\usepackage{xspace}
\usepackage{booktabs}
\usepackage{threeparttable}
\usepackage{adjustbox}

\DeclareUnicodeCharacter{0301}{\'{e}}

\usepackage{epstopdf}%This line makes .eps figures into .pdf - please comment out if not required.

\definecolor{cream}{RGB}{222,217,201}

\begin{document}

\pagestyle{fancy}
\thispagestyle{plain}
\fancypagestyle{plain}{
%%%HEADER%%%
\renewcommand{\headrulewidth}{0pt}
}
%%%END OF HEADER%%%

%%%PAGE SETUP - Please do not change any commands within this section%%%
\makeFNbottom
\makeatletter
\renewcommand\LARGE{\@setfontsize\LARGE{15pt}{17}}
\renewcommand\Large{\@setfontsize\Large{12pt}{14}}
\renewcommand\large{\@setfontsize\large{10pt}{12}}
\renewcommand\footnotesize{\@setfontsize\footnotesize{7pt}{10}}
\makeatother

\renewcommand{\thefootnote}{\fnsymbol{footnote}}
\renewcommand\footnoterule{\vspace*{1pt}% 
\color{cream}\hrule width 3.5in height 0.4pt \color{black}\vspace*{5pt}} 
\setcounter{secnumdepth}{5}

\makeatletter 
\renewcommand\@biblabel[1]{#1}            
\renewcommand\@makefntext[1]% 
{\noindent\makebox[0pt][r]{\@thefnmark\,}#1}
\makeatother 
\renewcommand{\figurename}{\small{Fig.}~}
\sectionfont{\sffamily\Large}
\subsectionfont{\normalsize}
\subsubsectionfont{\bf}
\setstretch{1.125} %In particular, please do not alter this line.
\setlength{\skip\footins}{0.8cm}
\setlength{\footnotesep}{0.25cm}
\setlength{\jot}{10pt}
\titlespacing*{\section}{0pt}{4pt}{4pt}
\titlespacing*{\subsection}{0pt}{15pt}{1pt}
%%%END OF PAGE SETUP%%%

%%%FOOTER%%%
\fancyfoot{}
\fancyfoot[LO,RE]{\vspace{-7.1pt}\includegraphics[height=9pt]{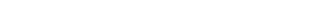}}
\fancyfoot[CO]{\vspace{-7.1pt}\hspace{13.2cm}\includegraphics{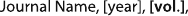}}
\fancyfoot[CE]{\vspace{-7.2pt}\hspace{-14.2cm}\includegraphics{head_foot/RF}}
\fancyfoot[RO]{\footnotesize{\sffamily{1--\pageref{LastPage} ~\textbar  \hspace{2pt}\thepage}}}
\fancyfoot[LE]{\footnotesize{\sffamily{\thepage~\textbar\hspace{3.45cm} 1--\pageref{LastPage}}}}
\fancyhead{}
\renewcommand{\headrulewidth}{0pt} 
\renewcommand{\footrulewidth}{0pt}
\setlength{\arrayrulewidth}{1pt}
\setlength{\columnsep}{6.5mm}
\setlength\bibsep{1pt}
%%%END OF FOOTER%%%

%%%FIGURE SETUP - please do not change any commands within this section%%%
\makeatletter 
\newlength{\figrulesep} 
\setlength{\figrulesep}{0.5\textfloatsep} 

\newcommand{\topfigrule}{\vspace*{-1pt}% 
\noindent{\color{cream}\rule[-\figrulesep]{\columnwidth}{1.5pt}} }

\newcommand{\botfigrule}{\vspace*{-2pt}% 
\noindent{\color{cream}\rule[\figrulesep]{\columnwidth}{1.5pt}} }

\newcommand{\dblfigrule}{\vspace*{-1pt}% 
\noindent{\color{cream}\rule[-\figrulesep]{\textwidth}{1.5pt}} }

\newcommand{\modelname}{DeltaDock\xspace}
\newcommand{\sitemodel}{LigPoc\xspace}

\makeatother
%%%END OF FIGURE SETUP%%%

%%%TITLE, AUTHORS AND ABSTRACT%%%
\twocolumn[
  \begin{@twocolumnfalse}
% {\includegraphics[height=30pt]{head_foot/journal_name}\hfill\raisebox{0pt}[0pt][0pt]{\includegraphics[height=55pt]{head_foot/RSC_LOGO_CMYK}}\\[1ex]
% \includegraphics[width=18.5cm]{head_foot/header_bar}}\par
% \vspace{1em}
% \sffamily
\begin{tabular}{m{4.5cm} p{13.5cm} }

\includegraphics{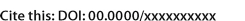} & \noindent\LARGE{\textbf{Multi-scale Iterative Refinement towards Robust and Versatile Molecular Docking}} \\%Article title goes here instead of the text "This is the title"
\vspace{0.3cm} & \vspace{0.3cm} \\

& \noindent\large{
Jiaxian Yan,\textit{$^{a}$} Zaixi Zhang,\textit{$^{a}$}} Kai Zhang,\textit{$^{a}$}
and Qi Liu$^{\ast}$\textit{$^{a}$}
\\

% & \noindent\large{Qi Liu$^{\ast}$\textit{$^{a\ddag}$}}
% \\

\includegraphics{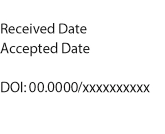} & \noindent\normalsize{Molecular docking is a key computational tool utilized to predict the binding conformations of small molecules to protein targets, which is fundamental in the design of novel drugs. Despite recent advancements in geometric deep learning-based approaches leading to significant improvements in blind docking efficiency, these methods have encountered notable challenges. These challenges include limited generalization performance on unseen proteins, the inability to concurrently address the settings of blind docking and site-specific docking, and the frequent occurrence of physical implausibilities such as inter-molecular steric clash. In this study, we introduce \modelname, a robust and versatile framework designed for efficient molecular docking to overcome these challenges.
\textbf{\modelname operates in a two-step process: rapid initial complex structures sampling followed by multi-scale iterative refinement of the initial structures.} In the initial stage, to sample accurate structures with high efficiency, we develop a ligand-dependent binding site prediction model founded on large protein models and graph neural networks. This model is then paired with GPU-accelerated sampling algorithms. 
The sampled structures are then updated using a multi-scale iterative refinement module that captures both protein-ligand atom-atom interactions and residue-atom interactions in the following stage.
Distinct from previous geometric deep learning methods that are strictly conditioned on the blind docking setting, \modelname demonstrates superior performance in both blind docking and site-specific docking settings. Comprehensive experimental results reveal that \modelname consistently surpasses all baseline methods in terms of docking accuracy. Furthermore, it displays remarkable generalization capabilities and proficiency for predicting physically valid structures, thereby attesting to its robustness and reliability in various scenarios.
}

\end{tabular}

 \end{@twocolumnfalse} \vspace{0.6cm}
 ]

\renewcommand*\rmdefault{bch}\normalfont\upshape
\rmfamily
\section*{}
\vspace{-1cm}

\footnotetext{\textit{$^{a}$~University of Science and Technology of China}, JinZhai Road, 230026, Anhui, China.}

\footnotetext{$\ast$~Corresponding author. qiliuql@ustc.edu.cn.}

%%%END OF FOOTNOTES%%%

%%%MAIN TEXT%%%%
\section{Introduction}
Deciphering the intricate structure of complexes formed between ligands and receptors remains a cornerstone in the realm of drug discovery~\cite{Boopathi2020Novel2C, Du2016InsightsIP}. The structural elucidation of these complexes enables researchers to undertake protein-ligand binding affinity predictions~\cite{Kitchen2004DockingAS, Jones2021ImprovedPB}, conduct comprehensive pharmacological analyses~\cite{Opo2021StructureBP}, and perform various other structure-oriented investigations, thereby facilitating the design of innovative and potent drug molecules. Recognizing its pivotal role, a multitude of computational tools have been devised to predict the structure of protein-ligand complexes. Among these, molecular docking stands as one of the most frequently utilized and impactful tools~\cite{Meng2011MolecularDA, deRuyck2016MolecularDA}.

Predominant molecular docking techniques such as VINA~\cite{Trott2010AutoDockVI}, SMINA~\cite{Koes2013LessonsLI}, and GLIDE~\cite{Friesner2004GlideAN} operate on a "sampling-and-scoring" paradigm to predict the structures of protein-ligand complexes, often referred to as binding conformations. In essence, these methodologies harness optimization algorithms like genetic algorithms~\cite{Goldberg1988GeneticAI} and simulated annealing~\cite{ Kirkpatrick1983OptimizationBS} to generate a suite of candidate binding conformations. The quality of these conformations is then evaluated using empirical or machine learning-based scoring functions~\cite{bohm1994development, hamelberg2004accelerated, gohlke2000knowledge, mooij2005general, yuriev2015improvements, Jain2006ScoringFF, Guedes2018EmpiricalSF, Li2020MachinelearningSF}.
Over the past four decades, these molecular docking methodologies have achieved significant successes~\cite{Beroza2022ChemicalSD, Gorgulla2020AnOD, Lyu2019UltralargeLD}. However, they are often marred by inefficiency and substantial computational resource requirements, primarily due to the vast number of structures sampled in pursuit of optimal binding conformations. For instance, docking 1 billion ligands using 10,000 CPU cores concurrently takes roughly two weeks~\cite{Gorgulla2020AnOD}, signifying a considerable expenditure of computational resources~\cite{Lyu2019UltralargeLD, Bender2021APG, Gloriam2019BiggerIB}. Consequently, there is a pressing need for molecular docking methodologies capable of accurately predicting binding conformations within the constraints of limited computational resources.

The recent surge in geometric deep learning~\cite{Bronstein2021GeometricDL} offers an innovative paradigm for molecular docking, proposing the direct prediction of binding conformations via geometric deep learning models. When contrasted with traditional molecular docking techniques, these geometric deep learning-based methods circumvent the need for intensive candidate sampling, thereby significantly enhancing docking efficiency~\cite{Strk2022EquiBindGD, Lu2022TANKBindTN}.
Since the pioneering work of Hannes et al.\cite{Strk2022EquiBindGD}, who first employed geometric deep learning models for molecular docking, a sequence of methods has been proposed, including TANKBIND\cite{Lu2022TANKBindTN}, E3BIND~\cite{Zhang2022E3BindAE}, and DiffDock~\cite{Corso2022DiffDockDS}. These methods primarily focus on blind docking scenarios (where the binding sites are unknown) and demonstrate significant performance advantages over traditional methods based on the root-mean-square deviation (RMSD) metric.
Nonetheless, despite these advancements, recent research has highlighted certain inherent limitations of such geometric methods. Their inadequate generalizability to novel proteins and a tendency to produce physically invalid poses have been particularly emphasized~\cite{Buttenschoen2023PoseBustersAD}. For instance, on the PDBbind dataset~\cite{Liu2015PDBwideCO}, EQUIBIND~\cite{Strk2022EquiBindGD} and TANKBIND~\cite{Lu2022TANKBindTN} only achieve docking success rates of 0.7\% and 4.9\% respectively when dealing with novel proteins that are absent from the training set. Notably, even the previously best-performing method, DiffDock, attains a success rate of 20.8\%, which is significantly lower than the 38.2\% success rate achieved when considering all test proteins, indicating a substantial performance drop-off. Moreover, poses predicted by geometric deep learning methods often exhibit physically implausible issues, such as internal steric clashes between protein-ligand and structural clashes within the ligand itself~\cite{Buttenschoen2023PoseBustersAD}. Furthermore, these methods primarily cater to blind docking scenarios and are unable to incorporate available pocket information in site-specific docking settings. These limitations pose significant barriers to the effective deployment of geometric deep learning-based molecular docking methodologies in practical applications.

\begin{figure*}[htp]
\includegraphics[width=\linewidth]{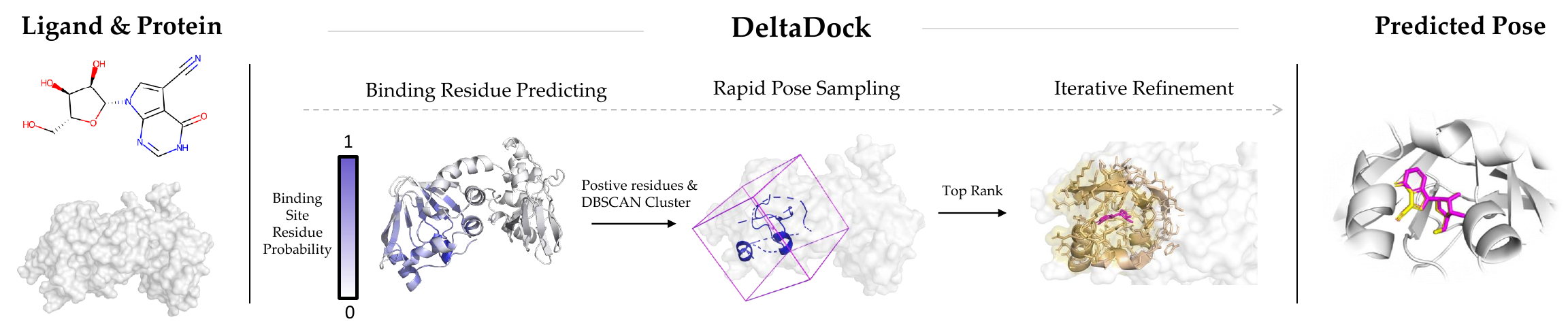}
\caption{
The framework of \modelname.
Left: The model takes as input the separate ligand and protein structures. 
Center: Three subprocesses of \modelname. Binding site residue probabilities are first predicted, and the positive residues are selected to perform DBSCAN. Using the center of DBSCN clusters, a fast pose sampling is performed. Finally, with the multi-scale iterative refinement, we update the top-ranked sampling pose.
Right: The final predicted pose. The yellow ligand structure is the ground-truth structure and the pink structure is the predicted structure.}
\label{fig:framework}
\end{figure*}

In order to address the above issues, we propose \modelname, a robust and versatile framework designed for efficient molecular docking.
Notably, \modelname operates through a two-step process to predict the ultimate binding structures. This process initiates with a swift sampling of initial structures, followed by a multi-scale iterative refinement of these structures (Fig.~\ref{fig:framework}).
(1)~During the initial sampling phase, we aim to sample accurate structures with high efficiency. To this end, we have developed a ligand-based binding site prediction model, \sitemodel, to narrow down the search space. \sitemodel leverages protein 1D sequence information from pre-trained large protein models (LPM) and 3D structure information extracted from trainable structure-aware equivariant graph neural networks. This multimodal information is amalgamated to generate accurate predictions. A GPU-accelerated sampling algorithm is then employed to sample structures within the predicted binding sites.
(2)~The subsequent phase involves the updating of sampled structures via a multi-scale iterative refinement module. Unlike previous work that only models the residues of proteins, thereby ignoring the fine-grained atom-atom interaction between protein side-chain and ligand, our refinement module is specially designed to capture both protein-ligand atom-atom interactions and residue-atom interactions. 
It is important to note that this two-step process can adeptly manage both blind docking and site-specific docking scenarios. In blind docking, \sitemodel is used to predict binding sites, whereas, in site-specific docking, the given binding sites are directly utilized.

To substantiate the efficacy of \modelname, we conducted extensive experiments to assess its prediction accuracy, efficiency, generalizability, and capacity to predict physically valid structures.
The experimental outcomes clearly show that \modelname consistently surpasses the baseline method across various scenarios in both blind docking and site-specific docking settings, while preserving high computational efficiency (approximately 1.96 seconds per protein-ligand pair).
Significantly, when confronted with new proteins unseen during training in the blind docking scenario, \modelname achieves a commendable 20.0\% improvement in the docking success rate compared to previous SOTA geometric deep learning-based methods, elevating it from 20.8\% to 40.1\%.
Moreover, a meticulous analysis of the physical properties of the predicted ligand structures reveals that \modelname effectively mitigates the issue of physically invalid conformations, thereby augmenting the overall reliability of the docking results.
In conclusion, \modelname demonstrates itself as a valuable and pragmatic approach, holding the potential to enhance our comprehension and implementation of molecular docking techniques. Its superior performance, efficiency, generalizability, and enhanced physical plausibility position it as a promising tool in the realm of drug discovery and design.

\section{Related Work}
\subsection{Sampling-based Docking Methods}
Traditional docking methods, epitomized by the likes of VINA~\cite{Trott2010AutoDockVI}, SMINA~\cite{Koes2013LessonsLI}, and QVINA~\cite{Friesner2004GlideAN}, are anchored in a "sampling-and-scoring" paradigm to find the optimal binding pose.
The efficiency and accuracy of these methods are swayed by the selection of scoring functions, search space, and sampling algorithms.
In the realm of site-specific docking, wherein binding sites are given,  the search space is defined as a cubic box that encompasses the binding sites, typically represented by a $22.5 \times 22.5 \times 22.5 $ \AA \; grid box~\cite{Trott2010AutoDockVI}.
In stark contrast, in blind docking, where the binding sites are unknown, the search space needs to cover the entire protein, resulting in a considerably larger search space.
To sample optimal poses within the defined search space, optimization algorithms like Broyden-Fletcher-Goldfarb-Shanno (BFGS)~\cite{Nocedal2000NumericalO} and Barzilai-Borwein (BB)~\cite{Qi2005OptimizationAC}, are employed to sample optimal poses in the defined search space on CPUs. This process typically involves a substantial number of steps (ranging from $10^4$ to $10^5$) and multiple copies (ranging from $8$ to $64$). Nonetheless, this laborious sampling sequence has not been immune to scholarly reproach, primarily due to its voracious computational demands.
Recently, some works have tried to accelerate the sampling process using GPUs, with notable mentions including Vina-GPU~\cite{Ding2023VinaGPU2F}, Uni-Dock~\cite{Yu2023UniDockGD}, and DSDP~\cite{Huang2023DSDPAB}. These methods use a large number of copies ($128-2048$) and short search steps ($20-200$) to capitalize fully on the parallel computational benefits proffered by GPUs.
This optimization strategy has proven to be highly effective, enabling the completion of a sampling process within a few seconds, which is approximately 10-fold faster in comparison to erstwhile CPU-oriented approaches.
% With such an optimization, it only takes about several seconds to finish a sampling process, $10 \times$ times faster than before.

\subsection{Geometric Deep Learning-based Docking Methods}
In recent years, an array of geometric deep learning-based docking methodologies has burgeoned, heralding a novel paradigm in the molecular docking arena. These methods aim to avoid the sampling process by formulating molecular docking as either a regression task or a generation task.
For instance, EQUIBIND~\cite{Strk2022EquiBindGD} utilizes an SE(3)-equivariant geometric deep learning model to facilitate direct-shot predictions by predicting a rotation and translation value to move the initial ligand structure. Concurrently, TANKBind~\cite{Lu2022TANKBindTN} predicts the inter-molecular distance and E3Bind~\cite{Zhang2022E3BindAE} directly predicts the ligand coordinates.
These methods typically treat the molecular docking task as a regression problem. 
In contrast to the prevalent approach, DiffDock~\cite{Corso2022DiffDockDS} pioneers an innovative ethos by framing it as a generative modeling problem and employing a diffusion model to generate a set of candidate poses for each input protein-ligand pair. A trained confidence model is then employed to pick out the most likely pose. Within the context of blind docking, DiffDock achieves significant performance improvement over sampling-based and geometric deep learning-based molecular docking methods.
Notwithstanding the exciting results achieved by geometric deep learning methods, recent research~\cite{Buttenschoen2023PoseBustersAD} highlights certain limitations, including poor generalization ability and physical invalidity. These limitations hinder the further development and widespread application of these methods. Efforts are needed to address these challenges and enhance the robustness and reliability of geometric deep learning-based docking approaches.

\subsection{Binding Site Prediction Methods}
As the foundation of structure-based drug design tasks such as molecular docking and \emph{de novo} ligand generation, the binding site prediction task has attracted expansive attention. A variety of methods have been developed for this task, encompassing traditional computational methods, machine learning methods, and geometric deep learning methods.
Traditional computational methods, exemplified by Fpocket~\cite{Guilloux2009FpocketAO} and CriticalFinder~\cite{Dias2017MultiGPUbasedDO}, employ geometric features, energy-related calculations, or similarity comparisons with libraries of known-function protein structures to identify binding sites.
Conversely, machine learning methods and geometric deep learning methods, such as P2rank~\cite{Krivk2018P2RankML}, PUResNet~\cite{Kandel2021PUResNetPO}, MaSIF~\cite{Gainza2019DecipheringIF}, and DeepSurf~\cite{Mylonas2020DeepSurfAS}, utilize traditional machine learning algorithms or geometric deep learning models to analyze and model the underlying patterns that correlate input data with the desired learning target.
These methods generally adopt ligand-independent approaches and focus on predicting all potential binding sites within individual proteins. By leveraging diverse models and techniques, binding site prediction methods contribute to the identification and characterization of binding sites, thereby facilitating subsequent tasks in the drug design process.

\section{Methods}
\label{sec:methods}
In this section, we introduce \modelname, our method specifically designed for molecular docking. We detail the steps for initial structure sampling and multi-scale iterative refinement in Sec.~\ref{sec:initial_sample} and Sec.~\ref{sec:iterative_refine}, respectively.
\begin{figure*}[htp]
  \includegraphics[width=\linewidth]{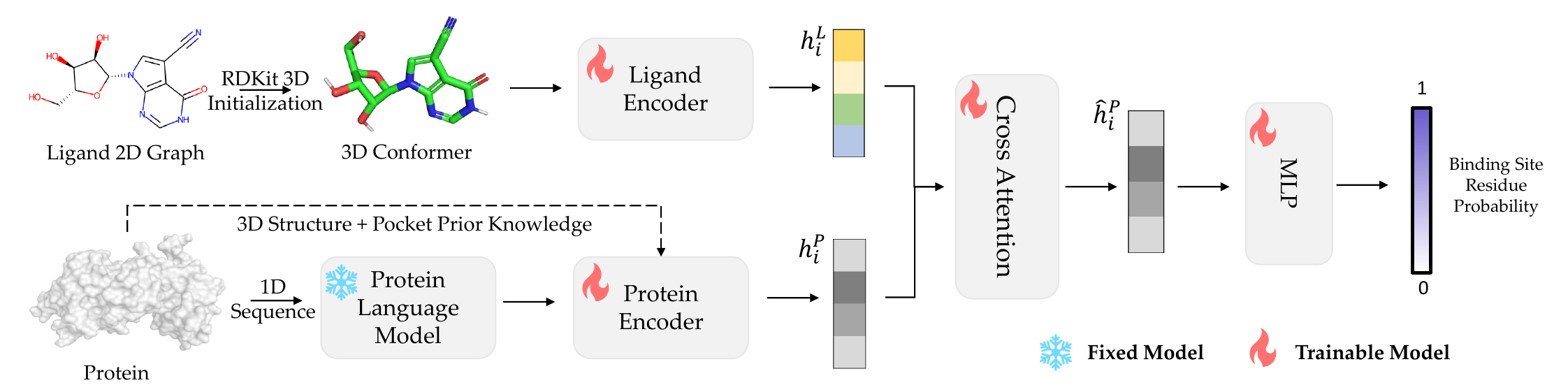}
  \caption{
  The framework of \sitemodel.
  The model takes as input the separate ligand and protein structures. Then a protein GNN encoder and a ligand GNN encoder are employed to extract informative node representations. A cross-attention mechanism is further applied to mix protein information and ligand information to make ligand-dependent binding site predictions.
  }
  \label{fig:ligpoc}
 \end{figure*}

\subsection{Initial Structure Sampling}
\label{sec:initial_sample} 
As mentioned before, \modelname initiates with a rapid sampling of initial complex structures. To narrow down the search space in the blind docking setting, we propose a ligand-based binding site prediction model, \sitemodel (detailed in Sec.~\ref{sec:ligpoc}). Subsequently, a GPU-accelerated sampling algorithm is employed to sample structures within the binding sites (discussed in~Sec.~\ref{sec:gpu_sample}).

\subsubsection{\sitemodel}
\label{sec:ligpoc}
Conceptually, given a protein $P$ and a ligand $L$, the target of \sitemodel is to predict the probability of each protein residue being a binding site residue and this can be formalized as:
\begin{equation}
    p_i = f(P, L), i \in \{1,2,...,N \},
\end{equation}
where $p_i$ denotes the probability of the $i$-th residue, $f$ is the model of \sitemodel, and $N$ is the total number of protein residues.
In this work, we define a residue as a binding site residue if its $\alpha$-carbon distance to any ligand atom is within $12.0$~\AA . 

The architecture of our binding site prediction model is depicted in Fig.~\ref{fig:ligpoc}. Unlike previous models that exclusively depend on protein structures for prediction, \sitemodel leverages information from both the individual ligand and the protein. A cross-attention mechanism is then applied to fuse the two individual information to make predictions.
% In the following context of this section, we detail the framework of \sitemodel.

\textbf{Ligand Encoder}. 
To extract ligand information, the input ligand $L$ is first represented as a ligand graph $\mathcal{G}^L=(\mathcal{V}^L, \mathcal{E}^L)$, where $\mathcal{V}^L$ is the node set and node $i$ represents the $i$-th atom in the ligand. 
% In this work, RdKit~\cite{Landrum2021rdkitrdkit2} is employed to generate 3D initial conformer of the input ligand.
Each node $v^L_i$ is also associated with an atom coordinate $c^L_i$ retrieved from the individual ligand structure $L$ and an atom feature vector $x^L_i$.
The edge set $\mathcal{E}^L$ is constructed according to the spatial distances among atoms. More formally, the edge set is defined to be: 
\begin{equation}
     \mathcal{E}^L = \left\{(i,j): |c^L_i - c^L_j|^2< cut^{L}, \forall i, j \in \mathcal{V}^L\right\},
\end{equation}
where $cut^{L}$ is a distance threshold, and each edge $(i,j)\in \mathcal{E}^L$ is associated with an edge feature vector $e^L_{ij}$. The node and edge features are obtained by RDKit~\cite{Landrum2021rdkitrdkit2}. 
A structure-aware graph neural network (GNN) is then used to perform message passing between ligand atoms in the ligand graph to extract informative node representations. Here, the GNN we used is AttentiveFP~\cite{Xiong2020PushingTB}, and the feature extraction process can be formalized as:
\begin{equation}
        H^L = GNN(\mathcal{G}^L),
\end{equation}
where $H^L$ is the ligand embedding matrix of shape $|\mathcal{V}^L| \times d$. And the $i$-th row of $H^L$, denoted by $h^L_i$, represents the embedding of the $i$-th ligand atom.

\textbf{Protein Encoder}. 
Then for protein, \sitemodel operates in a similar way, representing the input protein $P$  as a protein graph  $\mathcal{G}^P=(\mathcal{V}^P, \mathcal{E}^P)$.
$\mathcal{V}^P$ is the node set and the node $i$ represents the $i$-th residue in the protein.
Each node $v^P_i$ is also associated with an $\alpha$-carbon coordinate of the $i$-th residue  $c^P_i$ retrieved from the individual protein structure and a residue feature vector $x^P_i$. 
The feature vector consolidates two distinct types of information: the residue information, which is extracted from the protein language model ESM2~\cite{lin2022language}, and the pocket prior knowledge, which is inferred from the ligand-independent pocket prediction method P2Rank~\cite{Krivk2018P2RankML}.
The edge set $\mathcal{E}^P$ is constructed according to the spatial distances among atoms. More formally, the edge set is defined to be:
 \begin{equation}
      \mathcal{E}^P = \left\{(i,j): |c^P_i - c^P_j|^2< cut^{P}, \forall i, j \in \mathcal{V}^P \right\},
 \end{equation}
 where $cut^{P}$ is a distance threshold, and each edge $(i,j)\in \mathcal{E}^P$ is associated with an edge feature vector $e^P_{ij}$.
 The edge features are obtained following~\cite{Corso2022DiffDockDS}.
 Another AttentiveFP is then used to perform message passing between protein residues to extract informative node representations, formally:
\begin{equation}
        H^P = GNN(\mathcal{G}^P),
\end{equation}
where $H^P$ is the protein embedding matrix of shape $|\mathcal{V}^P| \times d$. The $i$-th row of $H^P$, denoted by $h^P_i$, represents the embedding of the $i$-th protein residue.

\textbf{Cross Attention}.
After extracting informative node representation from ligand structure and protein structure, we employ the cross-attention mechanism to capture protein-ligand mutual interaction and predict possible binding site residues. The cross-attention layer can be formally written as:
\begin{align}
    Q = H^PW^Q, K = H^LW^K, V = H^LW^V, \\
    \hat{H}^P = Attention(Q, K, V) = softmax(\frac{QK^T}{\sqrt{d_k}})V,
\end{align}
where $W^Q$, $W^K$, and $W^V$ are the transformation metrixs, and $d_k$ is the dimension of these metrixs. The matrix $\hat{H}^P$ represents the cross-attention protein embedding. Each row within $\hat{H}^P$, denoted as $\hat{h}^P_i$, corresponds to the cross-attention embedding of the $i$-th residue in the protein.

A multi-layer perception (MLP) and a Sigmoid layer are then used to predict the probability:
\begin{equation}
    p_i = \sigma (MLP(\hat{h}^P_i)),
\end{equation}
where $\sigma$ is the Sigmoid layer.

\textbf{DBSCAN Clustering}.
To determine the binding sites with predicted probability, we finally employ a DBSCAN~\cite{Ester1996ADA} algorithm implemented in SciKit-learn~\cite{Pedregosa2011ScikitlearnML} to cluster the positive residues as predicted by the \sitemodel. If the count of these residues falls below 50, we select the top 50 residues based on their probability values as positive residues. The "min\_samples" parameter and "eps" parameter are set to 2 and 9.0 respectively. Other parameters are set as default values. 
In the \modelname, we define pockets using these binding site residues, and the geometric center of binding site residues is set as the center of the binding site.

\begin{figure*}[htp]
  \includegraphics[width=\linewidth]{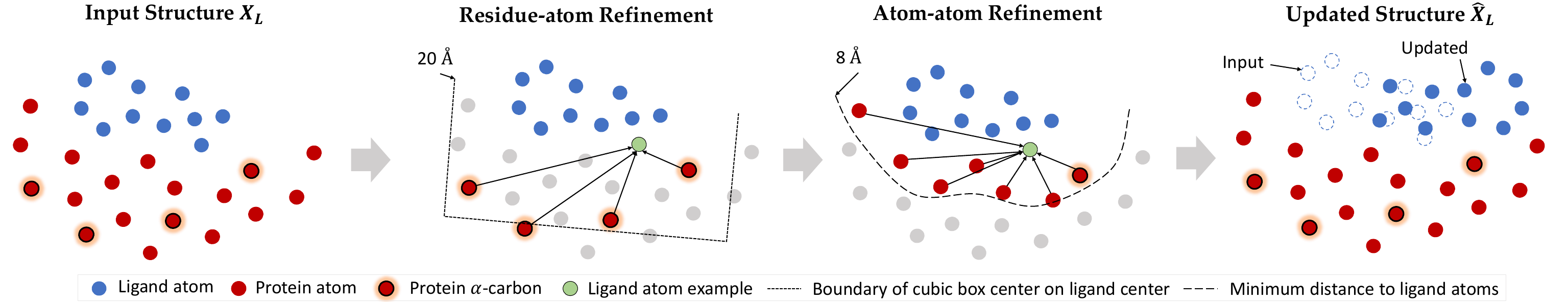}
  \caption{Overview of multi-scale iterative refinement module. Taking a complex structure as input, T rounds of residue-atom refinement and T rounds of atom-atom refinement are applied to update the input structure. For residue-atom refinement, the interaction between ligand atoms and protein residues within a $20.0 \times 20.0 \times 20.0$ \AA \; cubix box centered on the ligand geometric center is considered to update the structure. Subsequently, for atom-atom refinement, the interaction between ligand atoms and protein atoms within 8.0 \AA \; radius is considered instead. The green node serves as an example atom of the ligand, visualizing the protein-ligand interactions considered by E(3)-equivariant layers.}
  \label{fig:structure_finetune}
 \end{figure*}
 
\subsubsection{GPU-accelerated Sampling Algorithm}
\label{sec:gpu_sample}
Once the center of the binding site is determined, we utilize the docking method adapted by Huang~et~al.~\cite{Huang2023DSDPAB} for structure sampling.  The BFGS~\cite{Nocedal2018NumericalO} serves as the optimization method and the scoring function is identical to that of AutoDock Vina.
In this work, the search steps number and the search copy number are set to 40 and 384, respectively. 
For each binding site predicted, a $30 \times 30 \times 30 $ \AA \; cubic search space is defined. 
Occasionally, the DBSCAN may yield multiple clusters, leading to the identification of more than one potential binding site. The sampling process will be performed at each binding site to select the binding pose will the highest score.
The pose with the highest score is selected to perform further structure updation.
Compared to directly searching the space covering the whole protein, we just need to search the binding sites identified by \sitemodel.
Such an operation can effectively reduce the searching time and has been applied in previous work~\cite{Huang2023DSDPAB}.

\subsection{Multi-scale Iterative Refinement}
\label{sec:iterative_refine}
The top-1 ranked sampled complex structure is then input to this step.
Formally, we denote a protein-ligand complex structure as $C=(V_P, V_L, X_P, X_L)$, where $V_P$ and $V_L$ are the atom nodes of the protein and the ligand, and $X_P$ and $X_L$ are the corresponding coordinates of these nodes.
We aim to predict a structure that is more accurate than the input sampling structure in this step. This can be expressed as:
\begin{equation}
    \hat{X}_L = f(V_P, V_L, X_P, X_L)),
\end{equation}
where $\hat{X}_L$ is the coordinates of the update structure, and $f$ is the iterative refinement model. Essentially, this objective can be understood as predicting the difference $\Delta X_L$ between the input structure and the ground-truth structure, as shown in:
\begin{equation}
    \hat{X}_L = X_L + \Delta X_L.
\end{equation}
% $\alpha$-carbon atoms and their coordinates are specially denoted as $V_{(P,\alpha)}$ and $X_{(P,\alpha)}$. 

To achieve this, we refine the input structure through a multi-scale iterative refinement process. This process comprises $T$ rounds of residue-atom refinement to capture protein-ligand residue-atom interaction, followed by $T$ rounds of atom-atom refinement to capture protein-ligand atom-atom interaction, as illustrated in~(Fig.~\ref{fig:finetune}). 

\subsubsection{Residue-atom Refinement}
% Within each round of residue-atom refinement, an E(3)-Equivariant layer is employed to compute the interaction between protein residues and ligand atoms, which serves to update the coordinates of the ligand within the complex structure.
\textbf{Pocket Residue Selection}.
Historically, geometric deep-learning models have concentrated on directly predicting the binding pose using individual ligands and proteins. This approach necessitates the modeling of entire proteins to identify binding pockets and thus predict the binding structure within these pockets. However, in the \modelname framework, the binding sites are predetermined, thus only requiring the modeling of pocket structures. As such, during the residue-atom refinement rounds, only the protein residues within a $20.0 \times 20.0 \times 20.0 $ \AA \; cubic region centered at the geometric centers of ligands are considered. In this context, the full-atom structure of proteins is not considered, and the coordinates of the $\alpha$-carbon of residues are selected to represent the position of the entire residues.

\textbf{E(3)-equivariant layer}. 
Within each round $l$ of residue-atom refinement, an E(3)-equivariant layer is employed to calculate the interaction between protein residues and ligand atoms. More specifically, this layer adheres to the message-passing paradigm~\cite{Gilmer2017NeuralMP} and consists of four functions: intra-message function, inter-message function, aggregate function, and update function. 

The intra-message function works to extract messages $m_{i,j}$ and $\hat{m}_{i,j}$ between a node $i$ and its neighbor nodes $j$ from the same molecule graph. $m_{i,j}$ is later used for the updating of node features and $\hat{m}_{i,j}$ for the updating node coordinates.
This function can be formally written as :
 \begin{align}
    & d_{i,j}^{(l)} = ||x_i^{(l)} - x_j^{(l)}||, \; \forall (i,j) \in \mathcal{E}_P \cup \mathcal{E}_L,\\
    & m_{i,j} = \varphi_m(h_i^{(l)}, h_j^{(l)}, d_{i,j}^{(l)},), \; \forall (i,j) \in \mathcal{E}_P \cup \mathcal{E}_L,\\
    & \hat{m}_{i,j} = (x_i^{(l)} - x_j^{(l)}) \cdot \varphi_{\hat{m}}(m_{i,j}), \; \forall (i,j) \in \mathcal{E}_P \cup \mathcal{E}_L, 
 \end{align}
where $d_{i,j}^{(l)}$ is the relative distance between node $i$ and node $j$, and $\varphi$ is a MLP. 

The inter-message function works to extract messages $\mu_{i,j}$ and $\hat{\mu}_{i,j}$ between a node $i$ and it's neighbour nodes $j$ from the other molecule graphs: 
 \begin{align}
    & \mu_{i,j} = \varphi_\mu(h_i^{(l)}, h_j^{(l)}, d_{i,j}^{(l)}), \; \forall i \in \mathcal{V}_P, j \in \mathcal{V}_L \, or \ i \in \mathcal{V}_L, j \in \mathcal{V}_P, \\
    & \hat{\mu}_{i,j} =
    (x_i^{(l)} - x_j^{(l)}) \cdot \varphi_{\hat{\mu}}(\mu_{i,j}),
    \; \forall i \in \mathcal{V}_P, j \in \mathcal{V}_L \, or \ i \in \mathcal{V}_L, j \in \mathcal{V}_P.
 \end{align}
 
After extracting inter-message and intra-message, the aggregation function aggregates the neighbor messages of the node $i$: 
 \begin{align}
    & m_i = \sum_{j \in \mathcal{N}(i)}m_{i,j}, \forall i \in \mathcal{V}_P \cup \mathcal{V}_L,\\
    & \mu_{i} = \sum_{j \in \mathcal{N}_*^{(l)}(i)} \varphi(\mu_{i,j}) \cdot \mu_{i,j}, \forall i \in \mathcal{V}_P \cup \mathcal{V}_L,\\
    & \hat{m}_i = \sum_{j \in \mathcal{N}(i) }\frac{1}{d_{i,j}^{(l)} + 1} \cdot \hat{m}_{i,j}, \\
    & \hat{\mu}_i = \sum_{j \in \mathcal{N}_*^{(l)}(i) }\frac{1}{d_{i,j}^{(l)} + 1} \cdot \hat{\mu}_{i,j},
 \end{align}
where $\mathcal{N}(i)$ is the neighbor of node $i$ in the same graph, and $\mathcal{N}_*^{(l)}(i) $ is the set of nodes associated with node $i$ in the other graph.

Finally, the update function updates the position and features of each node: 
 \begin{align}
    & x_i^{(l+1)}
    = \eta x_i^{(0)} + (1 - \eta) x_i^{(l)} + \hat{m}_i + \hat{\mu}_i, \; \forall i \in \mathcal{V}_P \cup \mathcal{V}_L, \\
    & h_i^{(l+1)} = (1 - \beta) \cdot h_i^{(l)} + \beta \cdot \varphi(h_i^{(l)}, m_i, \mu_i, h_i^{(0)}), \; \forall i \in \mathcal{V}_P \cup \mathcal{V}_L,
\end{align}
where $\beta$ and $\eta$ are feature skip connection weight and coordinates skip connection weight, respectively. Through such a message-passing paradigm, our E(3)-equivariant layers make to update coordinates iteratively.

\subsubsection{Atom-atom Refinement}
The structures refined through $T$ rounds of residue-atom refinement are subsequently used as input for the atom-atom refinement rounds. In these rounds, protein atoms within an $8.0$ \AA \; radius of the ligand atoms are considered. In a manner analogous to the residue-atom refinement rounds, E(3)-equivariant layers are deployed to calculate the interactions between protein and ligand atoms, thereby facilitating iterative updates of the structure.

Fundamentally, the primary distinction between the residue-atom refinement rounds and the atom-atom refinement rounds lies in the protein information they model.
Given the considerable computational cost associated with directly modeling the full-atom structure of proteins, our design aims to mitigate excessive resource consumption. The multi-scale iterative refinement process not only allows the model to encompass protein residues across a relatively expansive spatial range (20 \AA), but it also fosters a detailed understanding of the interactions between ligand and protein atoms within a more focused range (8 \AA).

\subsubsection{Fast Structure Correction}
\textbf{Torsion Alignment}.
As the multi-scale iterative refinement layers update structures by modifying the coordinates rather than the torsional angles, as is done in methods like DiffDock and other sampling-based methods, it is crucial to ensure the plausibility of bond lengths and bond angles of the updated structure $\hat{X}_L$.
To this end, we employ a rapid torsion alignment for the updated structure.
The target of this alignment is to align the input structure $X_L$ with the updated structures $\hat{X}_L$ by rotating its torsional bonds.
Formally, let $(b_i, c_i)$ denote a $i$-th rotatable bond, where $b_i$ and $c_i$ are the starting and ending atoms of the bond, respectively. 
We randomly select a neighboring atom $a_i$ of $b_i$ and a neighbor atom $d_i$ of $c_i$ to calculate the dihedral angle $\hat{\delta}_i = \angle(a_ib_ic_i,b_ic_id_i)$ based on updated structure coordinates $\hat{X}_L$. Subsequently, we rotate the rotatable bond $(b_i, c_i)$ of input structures to match its dihedral angle $\delta_i$ the same as $\hat{\delta}_i$. This simple operation can be implemented efficiently using RDKit. 
After all rotatable bonds have been rotated, we align the rotated input structure to the updated structures to obtain the torsionally aligned structure $\hat{X}_L^{A}$. This process ensures the plausibility of bond lengths and bond angles in the torsionally aligned structure $\hat{X}_L^{A}$.

\textbf{Energy Minimization}. 
To further enhance the reliability of \modelname, we implement an energy minimization on the torsionally aligned structure $\hat{X}_L^{A}$, when an inter-molecular steric clash between the protein and ligand is detected. This energy minimization is conducted using SMINA~\cite{Koes2013LessonsLI}, a highly efficient tool for this process.
In this work, we denote structure after energy minimization as $X^{E}_L$.

% On average, the minimization process requires approximately 0.02 seconds for each data point in the testing dataset, demonstrating its computational efficiency.

\subsection{Training Objects}
\sitemodel and multi-scale iterative refinement layers in \modelname are trained with different loss functions separately.
\subsubsection{\sitemodel}
For \sitemodel, binding site residues are defined as the residues within 12 \AA \; of the ligand atoms.
This is an imbalance binary classification task, where binding sites residue only occupies about 19.7\% in the training dataset.
Considering this problem, we employ a focal loss~\cite{Lin2017FocalLF}, which has been widely used in imbalance classification conditions, such as objection detection.
Formally, the cross-entropy loss for binary classification task can be written as:
\begin{equation}
    CE(p,y) = \begin{cases}
        -log(p), \; if \; y=1,\\
        -log(1-p), \; others,
    \end{cases}
\end{equation}
where $y$ is the target label and $p$ is the predicted probability. 

The focal loss can be written as:
\begin{equation}
    FL(p,y) = \begin{cases}
        - \alpha (1-p)^\gamma log(p), \; if \; y=1,\\
        - (1-\alpha)p^\gamma log(1-p), \; others,
    \end{cases}
\end{equation}
where $\alpha$ is the weighting factor to balance positive and negative examples and $\gamma$ is the modulating factor to balance easy and hard examples. The two factor works together to relieve the problem of imbalance classification.
In this work, we set $\alpha=0.25$ and $\gamma=2$, respectively.

\subsubsection{Multi-scale Iterative Refinement Layers}
We design a physics-informed loss function for the multi-scale iterative refinement layers. The updated structures, denoted as $\hat{X}_L$, are employed in the computation of this loss. Formally, the loss function can be expressed as follows:
\begin{align}
     & \mathcal{L} =
     \lambda_1 \mathcal{L}_{intra} +
     \lambda_2 \mathcal{L}_{inter} +
     \lambda_3 \mathcal{L}_{bond}, \\
     & \mathcal{L}_{intra} =
     \frac{1}{N^2} \sum_{i \in \mathcal{V}_L} \sum_{j \in \mathcal{V}_L} |\hat{d}_{i,j} - d_{i,j}^*|, \\
     & \mathcal{L}_{inter} =
     \frac{1}{|\mathcal{E}_{pocket}|} \sum_{(i,j) \in \mathcal{E}_{pocket}}  |\hat{d}_{i,j} - d_{i,j}^*|, \\
     & \mathcal{L}_{bond} = 
     \frac{1}{|\mathcal{E}_{bond}|} \sum_{(i,j) \in \mathcal{E}_{bond}}  |\hat{d}_{i,j} - d_{i,j}^*|, 
\end{align}
where $\mathcal{L}_{inter}$ is the error between the predicted distance matrix between the protein binding site nodes and the ligand nodes and the ground truth distance matrix. This term is used to force the model to predict correct protein-ligand relative distance and thus predict accurate binding poses.
$\mathcal{L}_{intra}$ is the error between the pairwise distance matrices of the predicted molecule and the ground truth pairwise distance matrices, which play a role in maintaining the molecular structure.
$\mathcal{L}_{bond}$ is the error between bond lengths of generated molecules and ground truth molecules. We use this term to force the model to maintain accurate bond lengths, which is important to physical plausibly.
% As we can observe, only relative distances between nodes are considered to calculate the loss, which is E(3)-Invaraint

 \begin{table*}[t]
 \centering
 \caption{Blind docking. All methods take unbound ligand structures (generated by RDKit) and unbound protein structures as input, trying to predict bound complex structures. The test set is composed of 363 protein-ligand structures crystallized after 2019 and curated by the PDBbind database. \modelname-U refers to the model variant that generates structures without implementing torsion alignment and energy minimization. The best results are {\bf bold}, and the second best results are \underline{underlined}.}
 \label{tbl:blind_docking_test_set}
\begin{adjustbox}{width=1\textwidth}
\begin{threeparttable}
\begin{tabular}{lccccccccccccc}
\toprule
\multirow{3}{*}{Method} &
\multirow{2}{*}{Time average} & \multicolumn{6}{c}{Ligand RMSD} & \multicolumn{6}{c}{Centroid Distance}\\
% \cline{3-14}
& & \multicolumn{4}{c}{Percentiles} & \multicolumn{2}{c}{ \% below Threshold}  & \multicolumn{4}{c}{Percentiles} & \multicolumn{2}{c}{ \% below Threshold} \\
% \cline{2-14}
& Seconds & 25\% & 50\% & 75\% & mean & 2\AA  & 5\AA & 25\% & 50\% & 75\% & mean & 2\AA & 5\AA \\
\midrule
QVINA-W
& 49 & 2.5 & 7.7 & 23.7 & 13.6 & 20.9 & 40.2 & 0.9 & 3.7 & 22.9 & 11.9 & 41.0 & 54.6 \\
GNINA
& 393 & 2.8 & 8.7 & 21.2 & 13.3 & 21.2 & 37.1 & 1.0 & 4.5 & 21.2 & 11.5 & 36.0 & 52.0 \\
VINA
& 119 & 3.7 & 6.8 & 14.6 & 10.1 & 10.3 & 36.2 & 1.5 & 4.1 & 13.3 & 8.2 & 32.3 & 55.2 \\ 
SMINA
& 146 & 3.8 & 8.1 & 13.5 & 12.1 & 13.5 & 33.9 & 1.3 & 3.7 & 16.2 & 9.8 & 38.0 & 55.9 \\
GLIDE
& 1405 & 2.6 & 9.3 & 21.8 & 16.2 & 21.8 & 33.6 & 0.8 & 5.6 & 26.9 & 14.4 & 36.1 & 48.7 \\
DSDP 
& 1.22* & 1.0 & 3.0 & 7.9 & 7.2 & 42.4 & 59.8 & \textbf{0.3} & 1.0 & 4.8 & 4.9 & 60.3 & 75.8 \\ 
\midrule
EquiBind-U
& \textbf{0.14*} & 3.3 & 5.7 & 9.7 & 7.8 & 7.2 & 42.4 & 1.3 & 2.6 & 7.4 & 5.6 & 40.0 & 67.5 \\
EquiBind
& \underline{0.16*} & 3.8 & 6.2 & 10.3 & 8.2 & 5.5 & 39.1 & 1.3 & 2.6 & 7.4 & 5.6 & 40.0 & 67.5 \\
TANKBind
& 1.42* & 2.4 & 4.0 & 7.7 & 7.4 & 19.0 & 59.8 & 0.9 & 1.8 & 4.1 & 5.5 & 54.8 & 78.5 \\
DiffDock
& 40* & 1.4 & 3.3 & 7.3 & - & 38.2 & 63.2 & 0.5 & 1.2 & 3.2 & - & 64.5 & 80.5 \\
\midrule
\textbf{\modelname-U}
&
1.89* & \textbf{1.0} & \textbf{2.7} & \textbf{6.2} & \textbf{5.5} & \textbf{44.6} & \textbf{66.7} & \textbf{0.3} & \textbf{0.8} & \textbf{2.5} & \textbf{3.6} & \textbf{70.8} & \textbf{86.5}  \\
\textbf{\modelname}
&
1.96* & \textbf{1.0} & \underline{2.9} & \underline{6.7} & \underline{5.6} & \underline{43.3} & \underline{65.3} & \textbf{0.3} & \textbf{0.8} & \underline{2.8} & \underline{3.7} & \underline{68.6} & \underline{84.0}  \\
\bottomrule
\end{tabular}
\begin{tablenotes}
 \item[1] The time of consumption is denoted with * if it corresponds to GPU time; in the absence of this symbol, the time pertains to CPU time.
\end{tablenotes}
\end{threeparttable}
\end{adjustbox}

\end{table*}

 \begin{figure*}[h]
  \includegraphics[width=\linewidth]{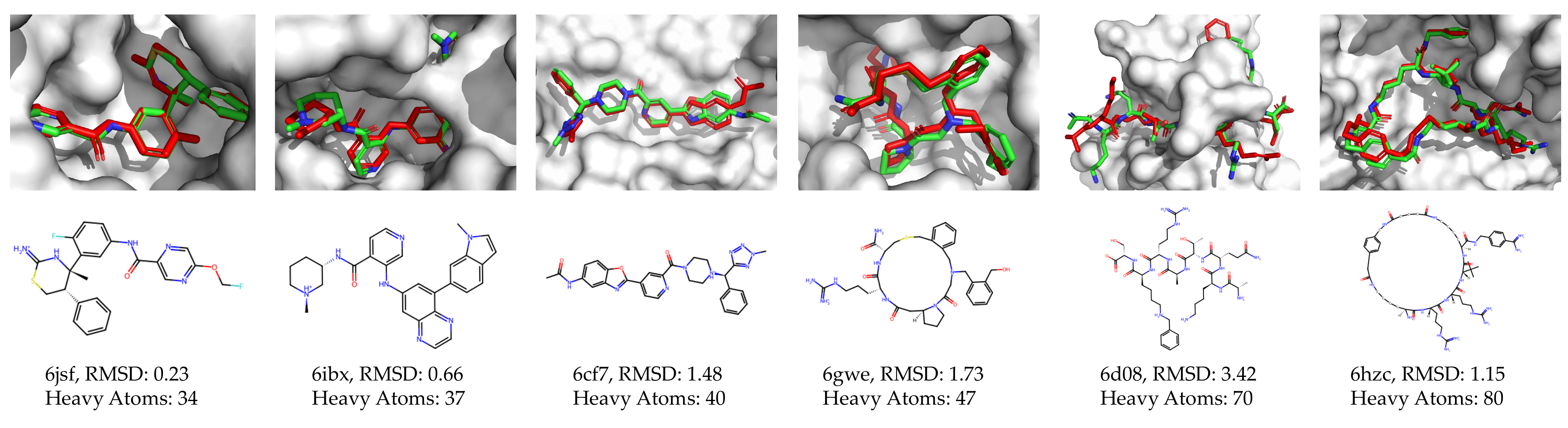}
  \caption{Examples for blind docking. Six examples with atom numbers ranging from $34$ to $80$ are selected. The green structures are predicted structures and the red structures are ground truth structures.}
  \label{fig:blind_dock_examples}
 \end{figure*}

\section{Results and Discussions}

\subsection{Baselines}
Mainstream geometric deep learning methods, EQUIBIND~\cite{Strk2022EquiBindGD}, TANKBIND~\cite{Lu2022TANKBindTN}, and DiffDock~\cite{Corso2022DiffDockDS}, and traditional sampling methods, VINA~\cite{Trott2010AutoDockVI}, SMINA~\cite{Koes2013LessonsLI}, and DSDP~\cite{Huang2023DSDPAB} are used as baselines. 
To make a fair comparison, we use the best results of these methods we can find in published papers.

\subsection{Metrics}
For the molecular docking task, root-mean-square-deviation (RMSD) and centroid distance are used to evaluate the docking accuracy of different docking methods, and the PoseBuster~\cite{Buttenschoen2023PoseBustersAD} test suite is employed to evaluate the performance of predicting physically valid poses.
% Then for the binding site prediction task, the distance between the center of the predicted pocket and the center of the ground-truth pocket (DCC) is employed. 

\subsection{Dataset}
\textbf{Test sets}. The \textbf{PDBbind}~\cite{Liu2017ForgingTB} v2020 dataset is the most commonly used dataset for the training and evaluation of geometric deep learning models.
We follow the time split strategy used in previous work~\cite{Strk2022EquiBindGD} to split the dataset to construct the train, validation, and test set.
All compounds discovered in or after 2019 are in the test and validation sets, and only those found before 2019 are in the training set.
The training set, validation set, and test set have 16,379, 968, and 363 complexes, respectively.
The overall performance of docking methods is evaluated on the time spit test set following previous works.
The generalization performance is tested in the PDBbind unseen test set, which contains 144 data with protein unseen in the training set, and the \textbf{PoseBuster} dataset containing 428 carefully selected data released from 1 January 2021 to 30 May 2023.
Finally, to evaluate the ability to predict physically valid poses, the PoseBuster test suite is performed on the PoseBuster dataset.
% In the test set, 144 data with protein unseen in the training set were specifically selected to form the test unseen set, which was employed to assess the model's generalizability. 

\textbf{Extended Training set}.
In addition to the training data from PDBbind v2020, we expanded our training dataset by filtering the Protein Data Bank (PDB)~\cite{Berman2003AnnouncingTW} as of August 2023.
The selection criteria incorporated the following: experimental resolutions lower than 4.0~\AA; ligands with more than 15 but fewer than 150 heavy atoms; successful RdKit sanitization; successful DSDP docking; deposited before 2019; absence of proteins in the unseen test set; and absence of ligands in both the test and validation sets. These rigorous data selection rules were implemented to expand our training dataset while minimizing the risk of data leakage into the test sets. As a result, we obtained an extended training set comprising 40,071 data points.

\subsection{Training Settings}
\subsubsection{\sitemodel}
The model was trained employing the Adam optimizer~\cite{Kingma2015AdamAM} with an initial learning rate of $10^{-3}$ and an $L_{2}$ regularization factor of $10^{-6}$.
The learning rate was scaled down by 0.6 if no drop in training loss was observed for 10 consecutive epochs.
The number of training epochs was set to 2000 with an early stopping rule of 200 epochs if no improvement in the validation performance was observed. 
As delineated in Sec.~\ref{sec:ligpoc}, the protein language model ESM2 is fixed and not incorporated into the training process.
In summation, the \sitemodel model is trained on two NVIDIA A100-PCIE-40GB GPUs for about 48 hours.

\subsubsection{Multi-scale Iterative Refinement Layers}
The Adam optimizer~\cite{Kingma2015AdamAM}, characterized by an initial learning rate of $10^{-3}$ and an $L_{2}$ regularization factor of $10^{-6}$, is employed for this process.
The learning rate was scaled down by 0.6 if no drop in training loss was observed for 10 consecutive epochs.
The number of training epochs was set to 1000 with an early stopping rule of 150 epochs if no improvement in the validation performance was observed. 
During the training of the multi-scale iterative refinement layers, two data augmentation methods were employed. Firstly, we applied binding site augmentation by adding noise in the range of 0.0 to 4.0 Å to the three dimensions of the predicted binding site center. This helped to enhance the robustness of the model by introducing variations in the binding site positions.
Secondly, we utilized input pose augmentation. Since the sampling methods generated multiple ligand poses per protein-ligand pair, we trained the model not only on the top-1 pose but also on the other top-10 poses. This allowed the model to learn from a diverse set of poses and be robust to the structures of input ligands.
When training, all residue-atom E(3)-equivariant layers and atom-atom E(3)-equivariant layers undergo joint training, with gradients back-propagating from the atom layer to the residue layer.
In total, these E(3)-equivariant layers were trained on two NVIDIA A100-PCIE-40GB GPUs for about 30 days. 

\subsection{Overall Performance on Molecular Docking}
In order to demonstrate the effectiveness and design validity, We first show the overall performance of the \modelname framework on the molecular docking task.
The blind docking setting and site-specific setting are tested both to evaluate the effectiveness and superiority of \modelname.

\subsubsection{Blind Docking}
\textbf{Docking accuracy.}
Blind docking, a scenario in which molecular docking is conducted without a predefined binding site, has consistently been a common goal among previous geometric deep learning methods.
Notably, \modelname has demonstrated its supremacy over all baseline methods, as illustrated in Table~\ref{tbl:blind_docking_test_set}. Specifically, \modelname achieved an impressive success rate of 43.3\% (where RMSD < 2.0 \AA), surpassing the previous state-of-the-art geometric deep learning method, DiffDock (38.2\%). It is worth highlighting that DiffDock operates as a generative model, necessitating 40 iterations of the diffusion generative process to sample 40 poses in order to attain the 38.2\% success rate. In stark contrast, \modelname functions intrinsically as a regression model, requiring a single inference.
Recent developments in GPU-accelerated docking methods, such as DSDP, have also made significant strides in blind docking. In comparison to DSDP, which stands as the top-performing sampling-based method within the PDBbind test set (42.4\%), \modelname consistently delivers superior results across all metrics, with a notable advantage in the centroid distance metric.
Interestingly, as expounded in Section~\ref{sec:gpu_sample}, \modelname utilizes the same sampling algorithm as DSDP. However, our framework enables \modelname to significantly surpass DSDP, thereby emphasizing the effectiveness of our approach.

\textbf{Docking Consuming Time.}
Efficiency, along with accuracy, is a critical performance metric in the evaluation of molecular docking methods.
As presented in Table~\ref{tbl:blind_docking_test_set}, despite the incorporation of a pre-trained protein language model (ESM2) and an energy minimization operation for enhanced accuracy and reliability, \modelname maintains a competitive level of efficiency, thanks to our meticulous design.
When compared to faster methods such as EquiBind, TANKBind, and DSDP, \modelname not only outperforms them in terms of docking accuracy but also demonstrates comparable efficiency within the same order of magnitude.
Moreover, \modelname surpasses traditional sampling-based methods such as VINA, SMINA, and GLIDE, as well as the previous state-of-the-art geometric deep learning method, DiffDock, by exhibiting significantly higher speed while simultaneously achieving improved docking accuracy.
Fundamentally, the development and application of molecular docking methods require a balance between efficiency and accuracy. The data presented in Table~\ref{tbl:blind_docking_test_set} indicate that \modelname could potentially serve as a viable tool for practical implementations.

  \begin{figure}[t]
  \includegraphics[width=\linewidth]{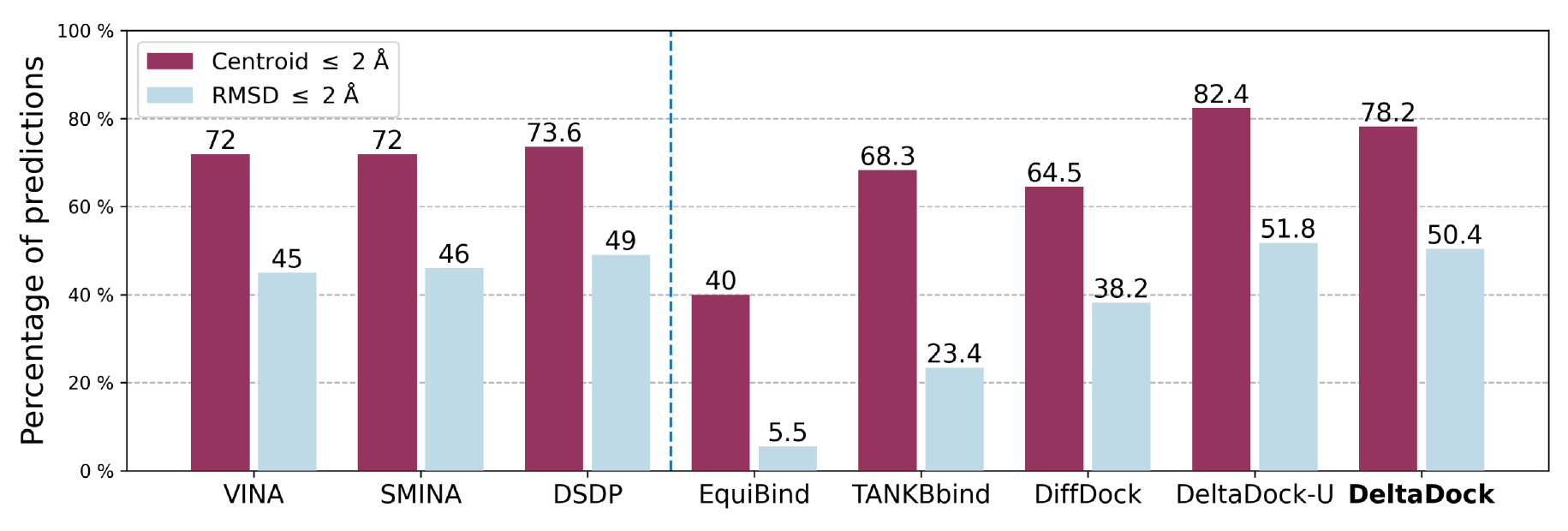}
  \caption{Site-specific docking. An Evaluation of the Performance of Various Molecular Docking Methods Utilizing the PDBbind Test Set. The search space was delineated by extending the minimum and maximum of the x, y, and z coordinates of the ligand by 4 \AA \; respectively. VINA, SMINA, and DSDP are traditional sampling-based methods, and other methods are geometric deep learning methods. }
  \label{fig:site_specific}
 \end{figure}

\subsubsection{Site-specific Docking}
Site-specific docking involves conducting molecular docking with a predefined binding site, a common scenario in real-world applications.
Most prior geometric deep learning methods, such as DiffDock and EquiBind, have predominantly focused on blind docking scenarios and do not readily adapt to site-specific docking requirements.
However, for \modelname, transitioning to the site-specific setting is straightforward. We simply replace the binding site prediction step with the use of known binding sites.
In this context, the site-specific docking performance of \modelname is depicted in Fig.\ref{fig:site_specific}.
For the geometric deep learning method TANKBind, which takes protein blocks as input, we directly supply the protein block with a radius of 20~\AA \; centered around the ground-truth ligand center to the model.
As for the sampling methods and \modelname, the search space is defined as a grid box centered at the geometric center of the ground-truth ligand structures. The dimensions of the grid box are determined by adding 4~\AA \; along the negative and positive directions for the minimum and maximum of the x, y, and z coordinates of the ligand, respectively.
Clearly, when provided with predefined binding sites, sampling methods yield significantly improved results. For instance, the docking success rate of VINA increased from 10.3\% to 45.0\%. In contrast, TANKBind only achieved a modest 4.4\% increase in success rate.
% This outcome suggests that, compared to sampling methods, previous geometric deep learning methods excel at identifying binding pockets rather than accurately predicting binding poses.
Nevertheless, \modelname continues to outperform all methods, demonstrating the adaptability of our framework to both blind docking and site-specific docking scenarios.

   \begin{table*}[h]
\centering
\caption{Blind docking for unseen receptors. All methods are evaluated on the new protein set, where the proteins have not been observed in the training set. The new protein set is composed of 142 protein-ligand structures crystallized after 2019 and curated by the PDBbind database. \modelname-U refers to the model variant that generates structures without implementing torsion alignment and energy minimization. The best results are {\bf bold}, and the second best results are \underline{underlined}.}
\label{tbl:blind_docking_unseen_set}
\begin{adjustbox}{width=1\textwidth}
\begin{threeparttable}
\begin{tabular}{lccccccccccccc}
\toprule
\multirow{3}{*}{Method} &
\multirow{2}{*}{Time average} & \multicolumn{6}{c}{Ligand RMSD} & \multicolumn{6}{c}{Centroid Distance}\\
% \cline{3-14}
& & \multicolumn{4}{c}{Percentiles} & \multicolumn{2}{c}{ \% below Threshold}  & \multicolumn{4}{c}{Percentiles} & \multicolumn{2}{c}{ \% below Threshold} \\
% \midrule{3-6}
& Seconds & 25\% & 50\% & 75\% & mean & 2\AA  & 5\AA & 25\% & 50\% & 75\% & mean & 2\AA & 5\AA \\
\midrule
QVINA-W
& 49 & 3.4 & 10.3 & 28.1 & 16.9 & 15.3 & 31.9 & 1.3 & 6.5 & 26.8 & 11.9 & 35.4 & 47.9 \\
GNINA
& 393 & 4.5 & 13.4 & 27.8 & 16.7 & 13.9 & 27.8 & 2.0 & 10.1 & 27.0 & 15.1 & 25.7 & 39.5 \\
VINA
& 119 & 5.0 & 9.6 & 19.0 & 12.8 & 7.8 & 25.5 & 2.2 & 6.1 & 17.8 & 10.9 & 24.1 & 41.8 \\
SMINA
& 130 & 4.8 & 10.9 & 26.0 & 15.7 & 9.0 & 25.7 & 1.6 & 6.5 & 25.7 & 13.6 & 29.9 & 41.7 \\
GLIDE
& 1405 & 3.4 & 18.0 & 31.4 & 19.6 & 19.6 & 28.7 & 1.1 & 17.6 & 29.1 & 18.1 & 29.4 & 40.6 \\
DSDP
& 1.22* & 1.0 & 4.5 & 10.0 & 9.1 & 37.2 & 54.9 & \textbf{0.2} & 1.5 & 6.5 & 7.5 & 54.2 & 69.0 \\
\midrule
EQUIBIND-U
& \textbf{0.14*} & 5.7 & 8.8 & 14.1 & 11.0 & 1.4 & 21.5 & 2.6 & 6.3 & 12.9 & 8.9 & 16.7 & 43.8 \\
EQUIBIND
& \underline{0.16*} & 5.9 & 9.1 & 14.3 & 11.3 & 0.7 & 18.8 & 2.6 & 6.3 & 12.9 & 8.9 & 16.7 & 43.8 \\
TANKBind
& 1.42* & 3.0 & 4.9 & 8.7 & 9.3 & 4.9 & 52.1 & 1.3 & 2.3 & 4.4 & 7.2 & 41.5 & 76.1 \\
DiffDock
& 40* & - & 6.2 & - & - & 20.8 & - & - & - & - & - & - & - \\
\midrule
\textbf{\modelname-U}
& 1.89* & \textbf{1.0} & \textbf{3.8} & \textbf{7.9} & \textbf{7.4} & \textbf{40.8} & \textbf{59.9} & \underline{0.3} & \textbf{1.0} & \textbf{4.2} & \textbf{5.6} & \textbf{64.8} & \textbf{77.5}\\
\textbf{\modelname}
& 1.96* & \textbf{1.0} & \underline{4.0} & \textbf{7.9} & \underline{7.6} & \underline{40.1} & \underline{58.5} & \underline{0.3} & \textbf{1.0} & \underline{4.6} & \underline{5.7} & \underline{62.0} & \underline{75.4}\\
\bottomrule
\end{tabular}
\begin{tablenotes}
 \item[1] The time of consumption is denoted with * if it corresponds to GPU time; in the absence of this symbol, the time pertains to CPU time.
\end{tablenotes}
\end{threeparttable}
\end{adjustbox}

\end{table*}

 \begin{figure*}[h]
  \includegraphics[width=\linewidth]{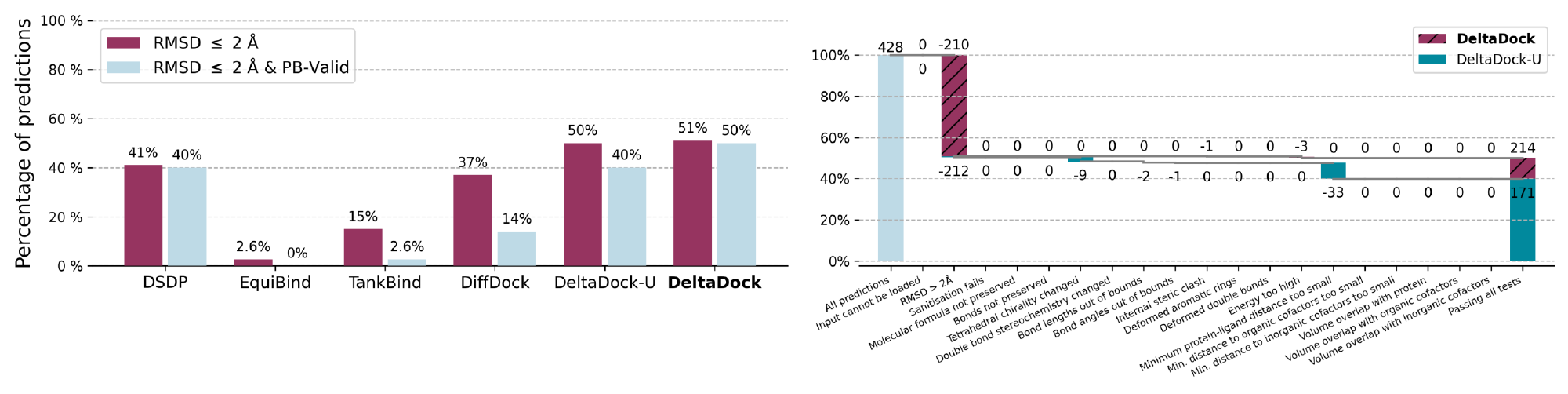}
  % \vspace{0.1cm}
  \caption{
  Blind docking performance on PoseBuster set.
  Left: Overall performance of different geometric deep learning methods.
  Right: A waterfall plot is provided, illustrating the PoseBuster tests as filters for both \modelname and \modelname-U predictions. The evaluation results for \modelname are denoted above the lines, while those for \modelname-U are annotated below the lines.}
  \label{fig:posebuster}
 \end{figure*}

   \begin{figure*}[h]
  \includegraphics[width=\linewidth]{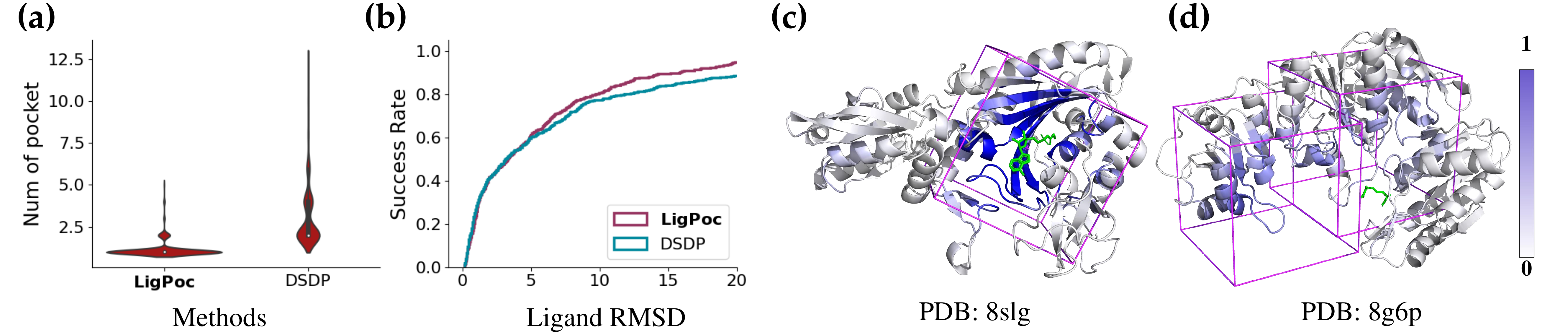}
  \caption{
  Performance of \sitemodel on the PoseBuster test set.
  (a) Pocket numbers violin plot of DSDP and \sitemodel.
  (b) RMSD cumulative curve of poses sampled on the binding sites predicted by DSDP and \sitemodel respectively.
  (c-d) Examples 8slg and 8g6p are given, demonstrating the binding site residue probability prediction and the determination of the search space for efficient sampling.
  In panels (c-d), protein residues predicted by \sitemodel to have a high probability of being binding site residues are represented with a deeper color.
  The ground-truth ligand structure is denoted in green.  As inferred from these illustrations, \sitemodel successfully predicts residues close to the ligand with high probability and thus determines accurate search spaces that cover the ground-truth ligand structure.}
  \label{fig:site_prediction}
 \end{figure*}
 
  \begin{figure}[htp]
  \includegraphics[width=1.0\linewidth]{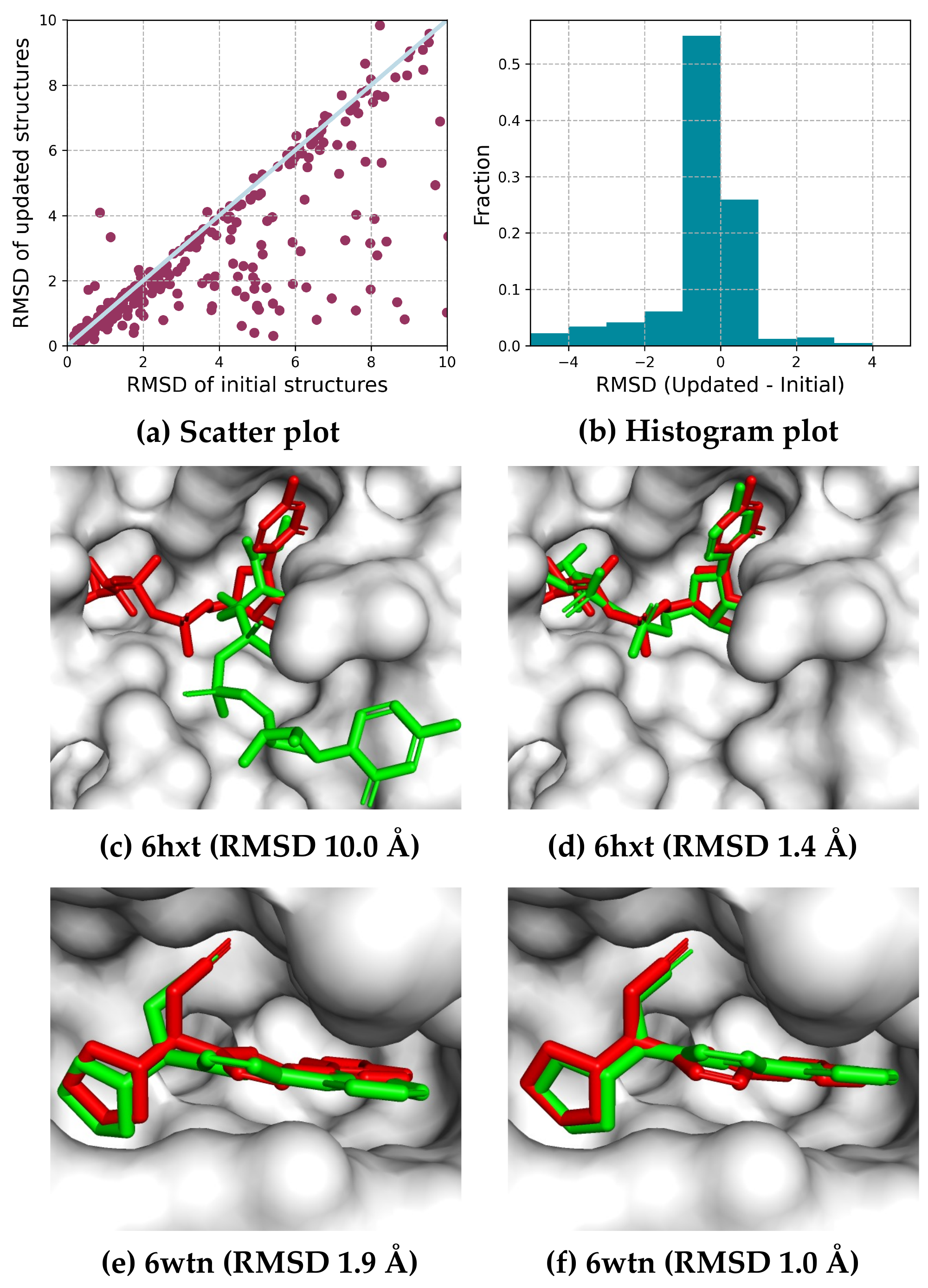}
  \caption{
  Performance of the multi-scale iterative refinement step on the PoseBuster test set.
  (a) Scatter plot of RMSD of initial structures and updated structures.
  (b) Histogram plot of RMSD changes after refining.
  (c-h) Examples of 6hxt and 6wtn. The left structures are input sampling structures and the right structures are updated structures. The green structures are predicted structures and the red structures are ground truth structures. }
  \label{fig:finetune}
 \end{figure}
\subsection{Evaluation of Generalization Capabilities and Prediction of Physically Valid Structures}
In this section, we delve into the capacity of \modelname to generalize effectively to previously unseen proteins and generate physically valid structures—an essential aspect for the practical application of molecular docking methods in real-world scenarios.

\subsubsection{Generalization Capabilities}
We first present the blind docking performance of \modelname and baseline methods on the PDBbind test unseen set following previous work.
Generally, as we can observe from Table~\ref{tbl:blind_docking_unseen_set}, the docking success rate of all methods on the PDBbind test unseen set is inferior to that on the PDBbind whole test set. 
For instance, the performance of GLIDE and QVINA-W experienced a modest decline of 2.2\% (21.8\% -> 19.6\%) and 5.6\% (20.9\% -> 15.3\%).
For previous geometric deep learning methods, the performance decrement was particularly conspicuous.
Remarkably, considering TANKBind and the SOTA method DiffDock as examples, they observed a performance drop of 14.1\% (19.0\% -> 4.9\%) and 17.4\% (38.2\% -> 20.8\%), respectively.
In the case of \modelname, it consistently manifests competitive performance, achieving a docking success rate of 40.1\%, thereby surpassing all other methods. \modelname outperforms the previous SOTA method, DiffDock, by an impressive 19.3\% in terms of success rate.

Subsequently, we extend our comparison of \modelname with prior methods to the PoseBuster set (Fig.~\ref{fig:posebuster}). For previous methods, their performance on this set surpasses that on the PDBbind test unseen set, albeit slightly inferior to that on the PDBbind whole test set.  While for \modelname, its performance surpasses that on the PDBbind whole test set, and significantly outstrips other methods. 
In summary, the performance of \modelname on these two distinct benchmark sets serves to substantiate its generalization capabilities. These results further amplify the exceptional generalization prowess of \modelname, thereby positioning it as a formidable contender in the domain of molecular docking, even when confronted with protein structures that have not been previously encountered on the training set.

\subsubsection{Prediction of Physically Valid Structures}
Then, we further analyze the ability of \modelname to predict physically valid structures by following the PoseBusters test suite designed by Buttenschoen et~el.~\cite{Buttenschoen2023PoseBustersAD}.
Besides the RMSD between predicted poses and ground-truth poses, 18 checks are included in the test suit, covering chemical validity and consistency, intramolecular validity, and intermolecular validity.
When physical validity is taken into account, the docking success rates of sampling methods remain robust, while the performance of previous geometric deep learning methods significantly deteriorates. 
As reported before~\cite{Buttenschoen2023PoseBustersAD}, different methods encounter different issues. For instance, out of 162 data points with an RMSD less than 2.0 \AA \; predicted by DiffDock, 90 instances exhibit issues related to too small distances between protein atoms and ligand atoms, indicating potential steric clashes between the protein and ligand.
For 64 data points with an RMSD less than 2.0 \AA \; predicted by TANKBind, the tetrahedral chirality of 17 instances changed after prediction, 19 instances had issues related to too small distances between protein atoms and ligand atoms, and only 11 instances passed all PoseBusters test suite checks.
Contrastingly, the \modelname model exhibited a near-perfect success rate, with almost 100\% of the data passing all PoseBusters test suite checks. This underscores the model's exceptional performance, particularly in terms of physical validity.
Moreover, the results provide a profound understanding of the impact of torsion alignment and energy minimization on the model's performance.
The \modelname-U variant, despite showing considerable improvement over previous methods even without the application of fast structure correction operations, still encounters minor issues. These include changes in tetrahedral chirality and instances where the protein-ligand distance is too small.
However, the implementation of fast structure correction operations largely resolves these issues in the \modelname model, demonstrating its superior performance and adaptability.
These results validate \modelname's ability to predict physically valid structures, thereby demonstrating its reliability for real-world applications.
 
\subsection{Further analysis of two steps in \modelname}
After the previous comparison, we proceed to further investigate the impact of various steps within the \modelname framework.

\subsubsection{\sitemodel}
\sitemodel model plays a pivotal role in \modelname. In theory, an accurate binding site prediction model facilitates the rapid pose sampling step to generate precise poses.
In this context, we compare \sitemodel with the previously established model DSDP~\cite{Huang2023DSDPAB}, which employs a 3D-CNN to make ligand-independent binding sites prediction.
We initially conducted a statistical examination of the number of binding sites predicted by these methods, and it is clear that \sitemodel predicts significantly fewer pockets than DSDP. As discussed in Section~\ref{sec:gpu_sample}, \modelname conducts site-specific docking at each predicted pocket, implying that a reduced number of pockets leads to accelerated inference speeds. While prior ligand-independent models are obliged to predict all potential pockets for all molecules across the chemical space, \sitemodel exclusively predicts potential pockets for target ligands, thereby abbreviating sampling time. This distinction emphasizes the advantage of our ligand-dependent model.
Additionally, we compare the poses sampled on the binding sites predicted by \sitemodel and DSDP. As illustrated in Fig.~\ref{fig:site_prediction}(b), \sitemodel, despite predicting fewer candidate pockets than DSDP, still demonstrates superior overall performance.
Lastly, in Fig.~\ref{fig:site_prediction}(c) and (d), we conduct a case study to visualize the prediction results of \sitemodel. It can be inferred from these figures that \sitemodel successfully predicts residues proximate to the ligand with high probability, thereby determining accurate search spaces that encompass the ground-truth ligand structure.
In summary, these findings underscore the accuracy and benefits of the \sitemodel.

\subsubsection{Multi-scale Iterative Refinement}
To evaluate the performance of the multi-scale iterative refinement module, we first assessed the ligand RMSD distribution before and after refinement on the PoseBuster dataset (Fig.~\ref{fig:finetune}).
For the majority of data points (309 out of 428), the refinement module effectively reduced the discrepancy between ground-truth structures and predicted structures. 
These data points initially exhibited RMSD values ranging from 0.2 \AA \; to 47.0 \AA \;, suggesting that the refinement step possesses a broad generalization capacity across different initial conditions.
While there was a deterioration in the quality of structures for 119 out of the 428 data points following the refinement process, it is important to highlight that the decrease in RMSD metrics in these instances is relatively minor when compared to the improvements observed in other structures. Only 13 data points' RMSD deteriorated by more than 1.0~\AA .

To further understand the impact of the refinement module, we conducted case studies in Fig.~\ref{fig:finetune}(c-f).
For example, in Fig.~\ref{fig:finetune}(c) and (d), we examined a scenario where the initial RMSD was relatively high, indicating significant discrepancies between the ground truth and predicted structures. The multi-scale iterative refinement led to a significant reduction in the RMSD value. This case underlines the model's ability to enhance prediction quality, particularly in instances of substantial initial discrepancies.
Then, Fig.~\ref{fig:finetune}(e) and (f) focus on a scenario where the initial RMSD was already quite low, indicating a reasonable alignment between the predicted and ground-truth structures. Despite the relatively low initial RMSD, the refinement module still managed to further refine the structure, illustrating that the model can further refine even already accurate predictions.
Collectively, these case studies demonstrate the effectiveness of the multi-scale iterative refinement module in enhancing the alignment between predicted protein-ligand structures and their ground-truth counterparts, regardless of the initial quality of the predictions.

\section{Conclusion}
Molecular docking serves as a critical instrument in the design of novel drugs. However, several challenges impede the advancement of geometric deep learning molecular docking, including the limited generalization performance on unseen proteins, the inability to concurrently address blind docking and site-specific docking scenarios, and the prevalence of physical implausibilities such as inter-molecular steric clashes.
To surmount these obstacles, we have introduced a robust and versatile framework, \modelname, designed for efficient molecular docking. \modelname operates through a two-step process that includes both an initial sampling phase and a subsequent iterative refinement of structures.
In the initial sampling phase, we aim to efficiently sample accurate structures. To this end, we have developed a ligand-based binding site prediction model, \sitemodel, which utilizes large protein models and graph neural networks to narrow down the search space. Following the prediction, a GPU-accelerated sampling algorithm is engaged to sample structures within the predicted binding sites.
In the subsequent phase, we employ a multi-scale iterative refinement module to update the sampled structures, aiming to predict more accurate binding conformations. 

In this paper, we have conducted comprehensive experiments to evaluate the prediction accuracy, efficiency, generalizability, and capacity of \modelname to predict physically valid structures. The results unambiguously demonstrate that \modelname consistently outperforms the baseline method across a variety of scenarios in both blind docking and site-specific docking settings, while maintaining high computational efficiency (approximately 0.8 seconds per protein-ligand pair).
Significantly, when tested with proteins unseen during training in the blind docking scenario, \modelname achieves a noteworthy 20.0\% improvement in the docking success rate compared to previous state-of-the-art geometric deep learning-based methods, raising it from 20.8\% to 40.1\%. Furthermore, a careful analysis of the physical properties of the predicted ligand structures reveals that \modelname effectively addresses the issue of physically invalid conformations, thereby enhancing the overall reliability of the docking results.
In conclusion, \modelname proves itself as a valuable and practical method, with the potential to augment our understanding and application of molecular docking techniques. Its superior performance, efficiency, generalizability, and enhanced physical plausibility position it as a promising tool in the field of drug discovery.

Looking forward, there are several potential extensions of \modelname that could serve as future works. For instance, we anticipate integrating an affinity prediction module into \modelname. Currently, \modelname does not include a module for predicting binding affinity as our primary focus has been on improving the quality of the predicted binding conformations. Additionally, although \modelname modeled the side-chain structures of proteins in this work, their flexibility is ignored. In future iterations, we aim to incorporate the aspect of side-chain flexibility.
Overall, we anticipate that further advancements in molecular docking models will significantly accelerate and benefit various applications.

\section*{Author Contributions}
Conceptualization: J.X.Y., Z.X.Z., and Q.L.;
Data Curation: J.X.Y.; 
Formal Analysis: J.X.Y. and Z.X.Z.;
Funding Acquisition: Q.L.;
Investigation: J.X.Y. and Z.X.Z.;
Methodology: J.X.Y., and Z.X.Z.;
Project administration: Q.L.;
Resources: Q.L.;
Software: J.X.Y.;
Supervision: Z.X.Z., K.Z., and Q.L.;
Validation: J.X.Y, K.Z., Z.X.Z., and Q.L.;
Visualization: J.X.Y, K.Z., and Z.X.Z.;
Writing – Original Draft: J.X.Y., K.Z., and Z.X.Z;
Writing – Review \& Editing: J.X.Y., Z.X.Z, K.Z., and Q.L..

\section*{Conflicts of interest}
There are no conflicts to declare.

\section*{Acknowledgements}
This research was supported by grants from the National Natural Science Foundation of China (Grants No. 61922073).
Our acknowledgment extends to the Anhui Province Key Laboratory of Big Data Analysis and Application (BDAA) laboratory server management team for their outstanding equipment and service support. 
We particularly express our sincere gratitude to Lixue Cheng for insightful discussions and valuable feedback.

%%%END OF MAIN TEXT%%%


\balance


%%%REFERENCES%%%
\bibliography{rsc} %You need to replace "rsc" on this line with the name of your .bib file
\bibliographystyle{rsc} %the RSC's .bst file

\end{document}